\documentclass[12pt]{article}

\usepackage[T1]{fontenc}
\usepackage[utf8]{inputenc}
\usepackage{multirow}
\usepackage{amsmath}
\usepackage{booktabs}
\usepackage{url}
\usepackage{graphicx}
\usepackage{amssymb}
\usepackage{authblk}

\newcounter{takeaway}[section]
\newenvironment{takeaway}[1][]{\refstepcounter{takeaway}\par\medskip
   \noindent \textbf{Takeaway~\thetakeaway. #1} }{\medskip}

\DeclareGraphicsExtensions{.pdf, .jpg, .png}
\graphicspath{{./}{./images/}}

\newcommand{\reffig}[1]{Fig.~\ref{#1}}
\newcommand{\reftab}[1]{Table~\ref{#1}}

\begin{document}

\title{
  Evaluating Endpoint-to-Path Cooperation in the Mobile Network using a 1-bit
  Rate-Delay Signal
}

\author[$\dagger$]{Thomas Fossati}
\author[$\star$]{Pedro A. Aranda Gutiérrez}
\author[$\ddag$]{Diego López}

\affil[$\dagger$]{Nokia, Cambridge, UK}
\affil[$\star$]{UC3M, Leganés, Spain}
\affil[$\ddag$]{Telefónica I+D, Madrid, Spain}

\maketitle

\begin{abstract}
This paper evaluates the impact of a Rate–Delay (RD) signal~\cite{PodlesnyG08}
-- specifically, the one described in~\cite{you-tsvwg-latency-loss-tradeoff-00}
-- on traffic crossing the mobile network through a set of experiments in a
simulated LTE network built with ns-3~\cite{ns3}.  In our experiments, we
compare a scenario with no RD treatment (i.e. a single best–effort Evolved
Packet System (EPS) bearer) with scenarios with RD treatment (i.e. separate EPS
bearers to carry RD–partitioned traffic) with honest and cheating users.  Our
objective is twofold:  On the one hand, we want to explore the suitability of
RD as a way to harmonize the Long–Term Evolution (LTE) and Internet Quality of
Service (QoS) models.  On the other hand, we aim at providing data to inform
protocol design as well as operations-related discussion on the notion of
exposing a 1-bit, clear-text RD signal from endpoints to the network path when
the use of end-to-end encrypted protocols would otherwise prevent flow
classification based on Deep Packet Inspection (DPI).  Specifically, the
question we want help answer is whether the gain in terms of end users' Quality
of Experience (QoE) and radio spectrum efficiency is enough to justify making
room for such signal.  All the experiments are fully documented and the tooling
used is made available as open-source to ensure their reproducibility.
\end{abstract}

\section{Introduction}
\label{S:1}

The amount of IP traffic flowing through the Mobile network is projected to
represent 20 percent of total IP traffic in the next five years, with 4G
amounting to more than three quarters of the share~\cite{cisco-forecast}.
Despite Long–Term Evolution (LTE) offering fine-grained Quality of Service
(QoS) control~\cite{3GPP-TS-23.203}, the typical 4G network at the time of this
writing is configured to carry all Internet-bound traffic over the same default
Evolved Packet System (EPS) bearer.  This means that all flows competing at the
bottleneck link -- which is usually located in the Radio Access Network (RAN)
-- are treated equally according to the QoS Class Identifier (QCI) parameters
associated with the default bearer, i.e. 300ms latency and $10^{-6}$ loss
budgets.  The latency bounds, in particular, are incompatible with interactive
applications e.g. Skype, Web based Real–Time Communication (WebRTC), online
gaming, as well as machine to machine applications (e.g., V2X),  that
critically require a low delay end-to-end path.  All these latency-sensitive
flows would probably benefit if isolated from other buffer filling flows,
especially when the bottleneck link is congested. When the induced queueing
delay is not properly bounded, the user experience of an interactive
application can become very frustrating.  Mobile Network Operators (MNOs)
over-provision their infrastructure to avoid congestion in the first
place~\cite{Martin-Geddes-2014}. But this strategy is clearly not sustainable,
as the spectrum is constrained by physics and cellular traffic is only destined
to grow in the coming years~\cite{cisco-forecast}.  The LTE QoS model is quite
rich, providing 15 different QCIs; so, what are the reasons pushing MNOs to
completely ignore it for Internet-bound traffic? Firstly, there is a cost
argument related to the maintenance of dedicated EPS bearers, including
configuration, increased control plane signaling to deal with the setup and
teardown operations, and run-time state that needs to be coordinated across a
variety of different LTE network nodes.  Secondly, the uncertainties associated
with mapping an incoming flow to the right EPS bearer, which include the trust
issue linked with QoS signaling at network interconnections: MNOs have no
incentives to honour QoS markings that are set by other administrative domains,
and possibly even by their own end-users~\cite{claffy2015}. In addition, flow
classification at line-rate is going to become much more difficult because of
the raise of encrypted and multiplexed transports (e.g.
QUIC~\cite{ietf-quic-transport}) which are, by definition, not amenable to deep
inspection.  Finally, the historical aversion of MNOs to favor Over-the-top
(OTT) real-time communication providers that are competing with their own
(MNOs-supplied) voice and video services and on their infrastructure. We note,
though, that this position is being re-evaluated thanks to the widespread
adoption of applications such as Facebook Messenger, Skype, Telegram or Viber
on smart phones.  If we manage to solve, or at least mitigate, the issues
listed above, we could grab the optimization opportunity that is hanging just
in front of us. Our intuition is that the Rate–Delay (RD)
approach~\cite{PodlesnyG08, you-tsvwg-latency-loss-tradeoff-00} might be the
silver bullet. In fact, RD addresses the cost argument by reducing the number
of dedicated bearers to only one: QCI 7 seems to be fit for purpose, having a
100ms delay upper-bound, at the cost of higher packet loss. RD also solves the
trust issues by creating a cooperative game between endpoints and the network
in which participants have no incentive to cheat. In fact, misrepresenting the
real nature of flows results in self-inflicted, unwanted, higher delay or loss
as a consequence of false signaling. Given the latter, the problem of efficient
flow classification could be solved by convincing encrypted traffic sources to
explicitly mark their traffic using a clear-text signal at a known position in
packets. If such signal was available, the mobile network could instantiate the
appropriate Traffic Flow Templates (TFTs) and dispatch at line-rate.  Finally,
the cost of extra state and control plane signaling needed by the dedicated low
latency bearer could be traded-off with the better utilisation of radio
resources that comes from the more relaxed Automatic Repeat Request (ARQ)
configuration required by the extra bearer to implement its shallow queue.

\subsection{Contribution}

In this paper, we verify that we can use QCI 7 to implement the RD  trade-off
by measuring the impact of using such strategy on end users' Quality of
Experience (QoE).  We describe and document a series of experiments conducted
using the ns-3  LTE module in order to compare the effect of RD markings on
traffic crossing a LTE network vs the best effort option, where no traffic
classification whatsoever exists. To the best of our knowledge, this is the
first time RD has been evaluated for use in mobile networks.

\subsection{Organization of the paper}

The rest of the paper is structured as follows: in Section~\ref{S:2} we
motivate the use of traffic classification using the RD trade-off; in
Section~\ref{S:3} we discuss the experimental setup used in the measurements
and, finally, in Section~\ref{S:4}we present our conclusions and discuss
further work.

\section{State of the Art}
\label{S:2}

End-to-end services in the Internet have been traditionally provided on a
best-effort basis. First approaches based on Integrated Services (IntServ), as
described in RFC 2210~\cite{rfc2210}, tried to signal QoS in an end-to-end path
by installing per-flow state in the network devices proved not to be scalable.
An alternative approach to provide better than best-effort in IP networks is
Differentiated Services (DiffServ) as described in RFC 2475~\cite{rfc2475}.
DiffServ can be used in a single domain to classify packets and make them
experience a specific Per–Hop Behaviour (PHB) that controls the QoE. Attempts
at providing scalable QoS management frameworks~\cite{mantar-diff-06} and
testbeds~\cite{qbone-isoc-99} were done during the late 1990's and early
2000's.  A full analysis of the technical and non-technical evolution of the
approach to implementing enhanced services in the Internet is provided
in~\cite{claffy2015}. Over-provisioning in IP networks and, specifically, in
mobile networks as a means to avoid congestion is documented
in~\cite{Martin-Geddes-2014}.  The migration of large volumes of traffic  from
clear-text to encrypted communications in the Internet is a confirmed trend, as
partially reflected in the increased use of encrypted Web
traffic~\cite{letsenc-stats}. Encryption, combined with flow
multiplexing,impacts the accuracy traditional traffic classification
methods~\cite{aceto2017}, which are normally implemented in network equipment
used by MNOs.  In an end-to-end encrypted scenario, the only reasonable place
to put QoS-related markings is in clear-text parts of the network or transport
layer protocols. In order to avoid man-in-the-middle attacks degrading the QoE,
it is desirable that these QoS-related markings are cryptographically
protected. Such a scenario, where marking and traffic classification to improve
QoE need user cooperation, calls for a framework that rewards honest and
penalizes dishonest use of QoS markings. One such approach is presented
in~\cite{PodlesnyG08}, where services are marked and treated depending on
whether they request high rates or low delay. High transmission rates imply the
presence of large queueing buffers to avoid loss, while low delay implies small
queueing buffers. A user requesting the wrong traffic class will be penalized,
because small queueing buffers are detrimental for high-rate traffic and long
queueing buffers will imply higher delay under high network utilization. This
approach was proposed in~\cite{you-tsvwg-latency-loss-tradeoff-00} as a
possible PHB for IP networks with DiffServ.

\section{Experiments}
\label{S:3}

\begin{figure}[t]
  \centering
  \includegraphics[width=.7\textwidth]{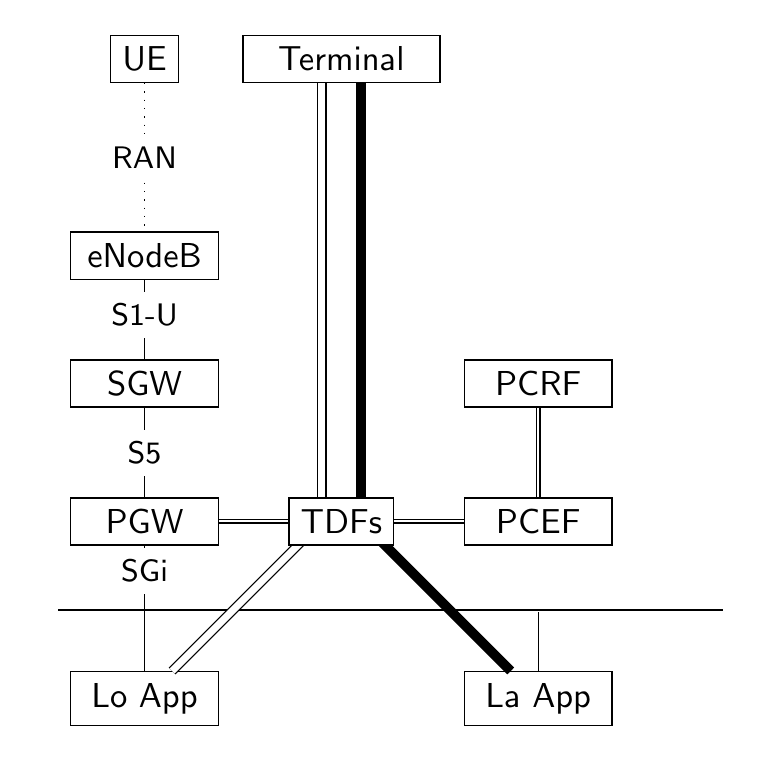}
  \caption{\label{fig:scenario2}Scenario used in the simulations}
\end{figure}

We use a number of ns-3 based simulations to measure the impact of
RD signalling coupled with a simplified LTE QoS setup that
uses one dedicated low-latency bearer in addition to the default bearer.

The crucial characteristic of this QoS framework is that cheating (i.e.,
using a certain marking to obtain an advantage over competing flows)
makes no sense to the end user because it may actually degrade their QoE.
This property is explicitly proven by one of our experiments.

In our implementation, we use the low-latency DiffServ codepoint (DSCP) defined
in~\cite{you-tsvwg-latency-loss-tradeoff-00}.  This choice makes it trivial to
specify the matching Traffic Flow Template (TFT) in the simulation environment.
It should be noted though that the use of the Loss-Latency Trade-off (LLT)
marking scheme in this context is just exemplary and, as long as the marker and
the classifier agree on the position and semantics of the signal used to
identify low latency flows, the specific kind of marking is irrelevant.

\reffig{fig:scenario2} shows the different elements and flows that are used in
the experiments.  The white flow represents a greedy TCP flow that is
not application limited (e.g., a large file download).  The black flow
represents a one way real-time flow with a bandwidth of 64kbps (e.g., a
real-time audio flow).  Marking, classification and the corresponding queuing
strategy depend on the scenario we test. For each set of QoS settings we
provide a control measurement where we apply the traffic to the network with
these settings deactivated and the experiment proper, where they are
activated.

\begin{table}[t]
  \centering
  \begin{tabular}{l l l}
    \toprule
    \multirow{2}{*}{LTE-uu} & FDD SISO, 6 RB & downlink peak: 4.4Mbit/s\\
    & baseline latency: &3ms\\ \midrule
    \multirow{2}{*}{S1/S5/S8} & Data rate: &5Mbit/s \\
    & propagation latency: &0ms \\ \midrule
    \multirow{2}{*}{SGi} & data rate: & 10Gbit/s \\
    & propagation latency: &1ms \\ \midrule
    eNB & proportional fair MAC scheduler&  \\ \midrule
    \multirow{2}{*}{Bearers} & default & QCI 9 \\
    & dedicated low-latency & QCI 7 \\
    \bottomrule
  \end{tabular}
  \caption{Configuration used in mobile simulations}
  \label{tab:lola-config}
\end{table}

The configuration of the SGi-LAN, LTE core and radio segments used in all
the experiments is described in Table~\ref{tab:lola-config}.

To evaluate the experiments, we use latency (delay and jitter) statistics.
In order to calculate the delay, we include a timestamp in the low-latency
packets and calculate the delay with the arrival timestamp. The jitter is
calculated from the set of measured delays ${d_1,..., d_N}$
using the following formula~\cite{jitter}:
\begin{equation*}
jitter = \frac{1}{N-1}\sum_{i=1}^{N-1} | d_{i+1} - d_i|
\end{equation*}

\subsubsection{Experiment 1: The Honest Marker}
\label{sec:res-honest}

In this experiment, we use two flows simulating a file download and a real-time
flow. We characterize the flows in the control with no marking: the mobile
network will put both on the same default bearer.  Then we enable LLT marking
on the real-time flow, which is routed through the dedicated low-latency bearer
as a consequence.  We simulate audio and video streams by controlling the size
of the packets and the rate at which they are generated.  In addition to
measuring the QoE parameters of the Constant Bit Rate (CBR) flow, we also
examine the throughput of the TCP flows.

\reftab{tab:honest-lat} compares the latency of the real-time flow in the
control scenario and the experiment. With the marking enabled a reduction by
72\% in the latency of the audio stream is achieved.

When we now examine the throughput of the TCP flow, \reftab{tab:honest-through}
shows that the measured throughput is slightly reduced when marking is used.
However, a variation of -0.39\% will likely have no impact on the experienced
QoE.

\begin{table}[t]
  \centering
  \label{tab:honest-lat}
  \begin{tabular}{l l c c c l }
    \toprule
    \multirow{2}{*}{Stream type} & \multirow{2}{*}{Run} & \multicolumn{3}{c}{Delay} & \multirow{2}{*}{Jitter} \\
    &  & mean & min & max &  \\
    \midrule
    \multirow{2}{*}{audio} & control & 15.48 & 5 & 24 & 4.43 \\
    & experiment & 4.32 & 4 & 6 & 0.72 \\
    \midrule
    \multirow{2}{*}{video} & control & 16.15 & 6 & 26 & 4.69 \\
    & experiment & 8.39 & 7 & 10 & 0.66 \\
    \bottomrule
  \end{tabular}
  \caption{Real-time flow latency (ms) in experiment 1}
\end{table}

\begin{table}[t]
  \centering
  \label{tab:honest-through}

  \begin{tabular}{l c }
    \toprule
    Run & Throughput \\
    \midrule
    no marking & 3.807 Mbit/s\\
    marking & 3.793 Mbit/s\\
    \bottomrule
  \end{tabular}
  \caption{TCP throughput in experiment 1}
\end{table}

\begin{takeaway}
  Improved delay and jitter for the real-time flow with negligible decrease in
  efficiency of the throughput seeking flow (and therefore of the RAN as
  a whole).
\end{takeaway}

\subsubsection{Experiment 2: The Cheater}
\label{sec:results-cheater}

In this scenario, we run two greedy TCP flows (large file downloads) and a
concurrent real-time flow.  We enable LLT marking on the real-time flow both in
the control and the experiment.  In the experiment, we also mark one of the
greedy TCP flows in order to route it through the low latency bearer together
with the real-time flow. This TCP flow simulates the cheater.

In this case, we use TCP retransmissions and the throughput to compare both
runs.  As shown in \reftab{tab:cheat-retrans} the cheater ends up
retransmitting a lot more (+460\%) which implies a substantial decrease in
throughput: The cheater gets -27.5\% throughput (honest gets a 18.85\% boost as
a consequence), as shown in \reftab{tab:cheat-throughput}.  Additionally, as
shown in \reftab{tab:cheater-cbr}, we observe that the real-time flow of the
cheater is not affected by the cheating TCP flow.

\begin{takeaway}
  Cheating penalizes the throughput of the greedy TCP flow significantly, while
  \textit{not} degrading the real-time flow.
\end{takeaway}

\begin{table}[h!]
  \centering
  \label{tab:cheat-retrans}
  \begin{tabular}{ l c c }
      \toprule
      Run & Honest & Liar \\
      \midrule
      no marking & 24 & 25\\
      marking & 31 & 140\\
      \bottomrule
  \end{tabular}
  \caption{Number of TCP retransmissions in experiment 2}
\end{table}

\begin{table}[h!]
  \centering
    \label{tab:cheat-throughput}
    \begin{tabular}{l  c c }
      \toprule
      Run & Honest & Liar \\
      \midrule
      no marking & 2.019 Mbit/s & 1.837 Mbits/s \\
      marking & 2.400 Mbit/s & 1.332 Mbits/s \\
      \bottomrule
    \end{tabular}
    \caption{TCP throughput in experiment 2}
\end{table}

\begin{table}[t!]
  \centering
  \label{tab:cheater-cbr}
  \begin{tabular}{l c c c c}
    \toprule
    Run & Mean & Min & Max & Stddev \\
    \midrule
    no marking & 5.373 & 4 & 9 & 1.156 \\
    marking & 5.373 & 4 & 9 & 1.156 \\
    \bottomrule
  \end{tabular}
  \caption{Influence on real--time latency (ms) of the cheater in experiment 2}
\end{table}

\subsubsection{Experiment 3: Multiple UE connected to one eNodeB}
\label{sec:manyusers-val}

In this experiment, we want to see how the system behaves when we have many
users together. For this experiment, we increase the bandwidth at the S1, S5
and S8 interfaces to 50 Mbps. We assume that marking users are \textit{honest}
users and examine the perceived QoE of marking and non-marking users as the
number of marking users increases in a constant population that produces an
aggregate bandwidth of latency-sensitive flows that does not surpass the
bandwidth available at the RAN (i.e., the bottleneck).

We worked with a population of 20 User Equipment (UE) per eNodeB. Each UE
receives a UDP stream simulating a video/audio stream and TCP traffic from a
greedy application running in the background.  In each run, we had a fraction
of the UE receive the CBR stream through the dedicated low latency bearer,
while the rest used the default bearer for both streams. The UDP streams were
started at random times uniformly distributed between 1 and 3~s, the streams
lasted for 10~s and the simulator ran for 14~s, assuring that the full CBR
stream was captured. We simulated an audio stream carried over RTP (over UDP)
with the following characteristics separately:

\begin{table}[h!]
  \centering
  \begin{tabular}{lccccc}
    \toprule
	Stream type & Packet size & PPS & effective BW & IP layer BW & \# of nodes\\
    \midrule
    audio & 160 & 50  & 64 &  80 & 20\\
    video & 100 & 400 & 320 & 352 & 10\\
    \bottomrule
  \end{tabular}
  \caption{\label{tab:stream}CBR stream parameters. Packet sizes are in bytes, bandwidth (BW) is in kbit/s}
\end{table}

\begin{figure}[t]
  \centering
  \includegraphics[page=1,width=.75\textwidth]{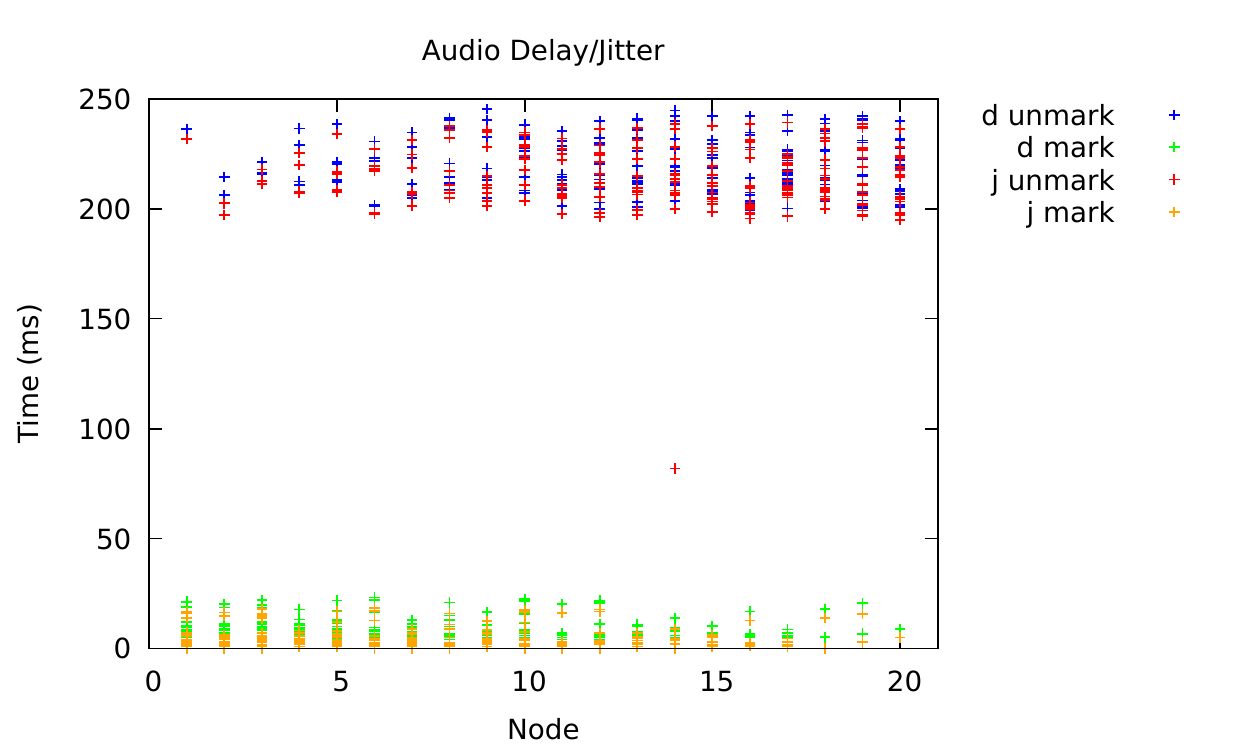}
  \caption{\label{fig:audio-simple}QoE parameters for 20 CBR flows @ 64kbit/s
  }
\end{figure}

\reffig{fig:audio-simple} shows the delay (d) and jitter (j) measured in each
of the UE. The measurements are grouped by colours depending on whether the
node received a marked or an unmarked audio stream. It can be clearly
appreciated that when marking was used (i.e., when QCI 7 was used for
latency-sensitive traffic) the nodes perceive significantly lower delay and
jitter. This implies that using LLT marking in these streams produced much
better QoE.  In particular, it is a well known fact~\cite{ITU-G.144} that the
quality of a voice call degrades rapidly where the mouth-to-ear delay latency
exceeds 200 ms.

\begin{figure}[t]
  \centering
  \includegraphics[page=1,angle=270,trim=40 40 40 40,width=.6\textwidth]{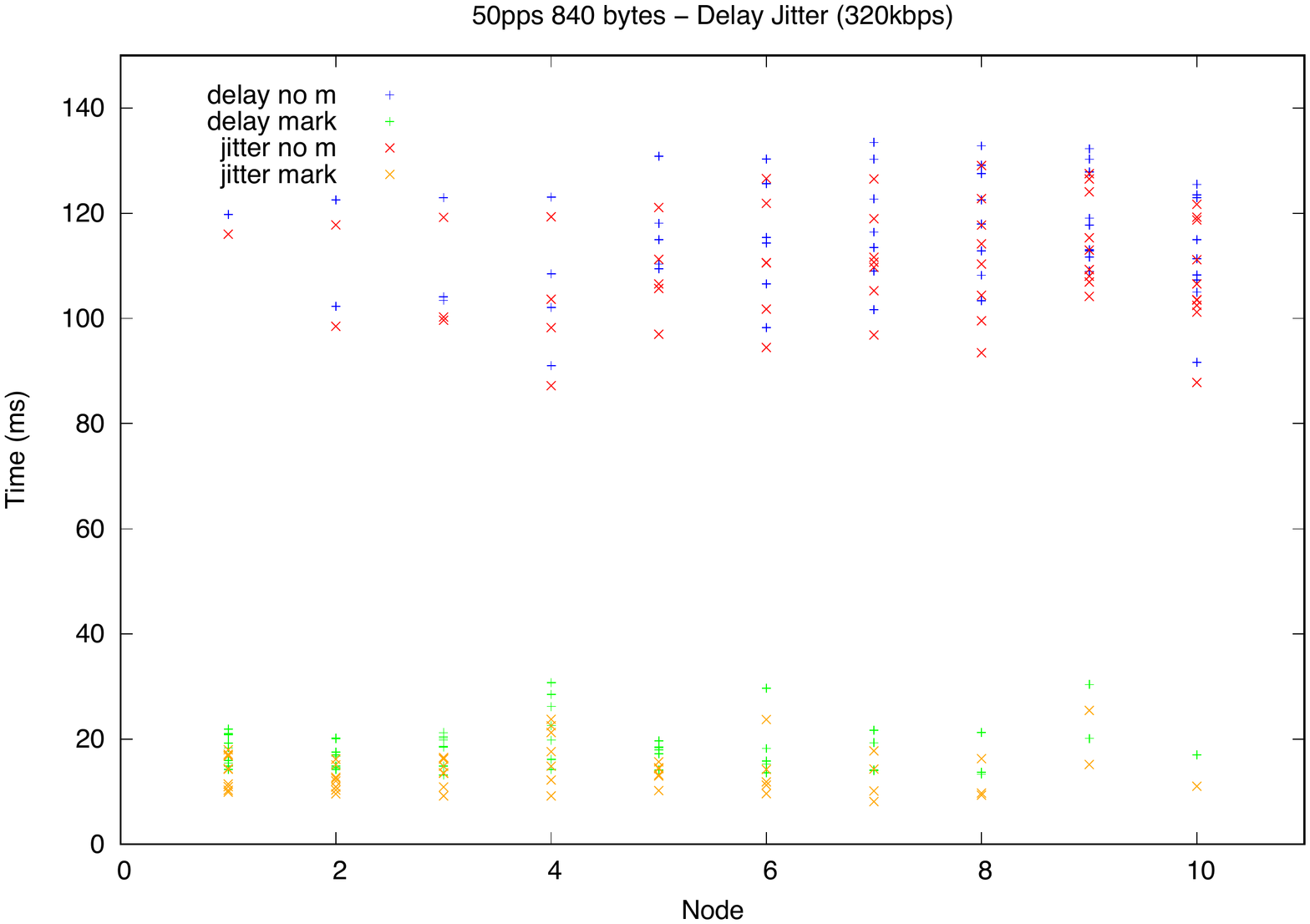}
  \caption{\label{fig:video-10}QoE parameters for 10 CBR flows @ 320kbit/s}
\end{figure}

We then repeated the experiment with 10 nodes and a 320kbit/s UDP CBR (video)
flow competing with a greedy TCP flow per node. We observe the same behaviour,
i.e., that marking yields better QoE behaviour than not marking. The results
are shown in \reffig{fig:video-10}.

\begin{takeaway}
  The dedicated low-latency bearer produces extremely good results, with delays
  one order of magnitude less than the default bearer.
\end{takeaway}

\section{Conclusions}
\label{S:4}

The simulations performed around low latency support mechanisms in mobile
networks show very promising results.  Moving the idea out of the testbed
requires first of all agreeing among the involved parties (mobile networks and
applications) on the format and semantics of the core signalling.  A mechanism
based on Non Queue Building Flows (NQB)~\cite{white-tsvwg-nqb-00} has been
recently proposed by the authors in the IETF TSV working
group~\cite{fossati-tsvwg-lola-00}.  This approach has the advantage of
harmonising the signalling between fixed and mobile access and at the same time
not precluding the possbility to further experiment with Low Latency Low Loss
Scalable throughput (L4S)~\cite{ietf-tsvwg-l4s-arch-03}.  This is the
groundwork needed for a few other standards and dissemination activities which
include at a minimum: recommendations for endpoint and API developers (for
example, IETF RTCWEB, W3C WebRTC, M2M-related SDOs), and guidelines for mobile
network operators -- ideally via GSMA.

\bibliographystyle{abbrv}
\small
\bibliography{lola}

\end{document}